\documentclass[floats,floatfix,showpacs,amssymb,prd,onecolumn,nofootinbib,twocolumn,groupedaddress]{revtex4-2}

\usepackage{graphicx}
\usepackage{dcolumn}
\usepackage{bm}
\usepackage{amsmath}
\usepackage[colorlinks=true,linkcolor=blue,citecolor=blue,urlcolor=blue]{hyperref}
\usepackage{dsfont}
\usepackage{physics}

\usepackage{booktabs}
\usepackage{threeparttable}
\usepackage{siunitx}
\usepackage{tablefootnote}

\usepackage{xcolor}
\usepackage{ulem}
\usepackage{makecell}

\newcommand{\nl}[1]{{\textcolor{blue}{{{#1}}}}}

\newcommand\comment[1]{}

\date{\today}

\begin{document}

\title{Constraints with CMB lensing on dark matter decays to massive decay products}
\author{Anna Bencke}
\email{abencke1@jh.edu}
\affiliation{William H. Miller III Department of Physics \& Astronomy, Johns Hopkins University,  3400 N.\ Charles St., Baltimore, Maryland 21218, USA}
\author{Nanoom Lee}
\affiliation{William H. Miller III Department of Physics \& Astronomy, Johns Hopkins University,  3400 N.\ Charles St., Baltimore, Maryland 21218, USA}
\author{Marc Kamionkowski}
\affiliation{William H. Miller III Department of Physics \& Astronomy, Johns Hopkins University, 3400 N.\ Charles St., Baltimore, Maryland 21218, USA}

\begin{abstract}
    Motivated by the recent measurements by the Dark Energy Spectroscopic Instrument (DESI) which suggest a late-time matter density approximately 5\% lower than that inferred from Planck we investigate models in which dark matter decays to two less massive, and thus warm, states. The decaying dark matter (DDM) models are parameterized by the fraction $f$ of dark matter that decays, the decay rate $\Gamma$, and the fraction $\varepsilon$ of the mass retained by the decay products. To efficiently explore the warm decay product regime, we employ \texttt{CLASSIER-DDM}, a modified version of the public Boltzmann solver \texttt{CLASS} that solves the perturbation equations with DDM via an integral-equation method. We consider DESI DR2 baryon acoustic oscillations and Planck 2018 CMB data including lensing, and find that DDM models are not favored over the $\Lambda$CDM model. This result arises because CMB lensing tightly constrains the velocity kicks imparted to decay products.  For example, for $f\simeq0.5$, we find $1\sigma$ constraints to the decay-product kick velocities an order of $10^{-2}$ to $10^{-3}$ times the speed of light for decay redshifts from shortly after recombination until today. Nonetheless, the allowed parameter space includes models with sufficient power suppression at small-scales to potentially address dwarf galaxy anomalies. Our results also suggest that explanations for DESI that involve dark-matter decays to one massive and one massless particle will be constrained by CMB lensing.
\end{abstract}

\maketitle

\section{Introduction}

The standard $\Lambda$CDM cosmological model continues to provide an excellent description of observations, from the Planck 2018 CMB measurements to large-scale structure surveys~\cite{Planck:2018vyg, DESI:2025zgx, DES:2017myr, Kosowsky:2003smi, Euclid:2024yrr, Casey:2022amu, SDSS:2000hjo}. However, recent results from the Dark Energy Spectroscopic Instrument (DESI) Data Release 2 have drawn attention to a potential tension, in that DESI BAO measurements yield a matter density  approximately 5\% lower than the Planck-inferred value~\cite{DESI:2025zgx, Planck:2018vyg}. This discrepancy admits a simple interpretation: the universe may contain less matter today than it did at recombination.

Decaying dark matter (DDM) provides a natural mechanism for such a reduction. If a fraction of the dark matter decays on cosmological timescales, the late-time matter density is diminished relative to its early-time value, potentially reconciling a Planck-like $\Omega_m$ at $z \sim 1100$ with a DESI-like $\Omega_m$ at $z \lesssim 2$. This scenario has been studied in the limit of massless decay products (``dark radiation''), where lifetimes are constrained to exceed the age of the universe~\cite{Enqvist:2019tsa, Nygaard:2020sow, Anchordoqui:2022gmw, Audren_2014, Poulin_2016, Lynch:2025}. More recent work has explored one dark radiation product and one warm (massive) decay product, which relaxes these constraints by reducing free-streaming effects~\cite{Aoyama:2011, Aoyama:2014tga, Abell_n_2021, Bucko:2022kss, Bucko:2024izb, Holm:2022eqq,Acharya:2026bmo}. To our knowledge, DDM with two massive decay products has received limited attention. This is in great part due to the numerical difficulty of evolving the Boltzmann hierarchy for non-cold species.

In this work, we study for the first time the evolution of cosmological perturbations with two warm decay products using \texttt{CLASSIER-DDM} \cite{Lee:2026zrl}, a modified version of the Boltzmann solver \texttt{CLASS} \cite{Lesgourgues:2011re} that solves the Boltzmann equations for warm relic perturbations efficiently via an integral-equation method~\cite{Ji:2022iji, Lee:2025zym, Lee:2025vgv}. This enables a comprehensive exploration of the DDM parameter space. We consider two-body decays into two decay products with equal masses $m_{\chi_{1,2}} = \varepsilon m_\chi /2$, parameterized by the unstable fraction $f$, the decay rate $\Gamma$, and the fraction $\varepsilon$ of the decaying-particle mass retained by the decay products, with $\varepsilon$ ranging from 0 (dark radiation) to 1 (CDM). 

Despite our original motivation, we find that models which accommodate the lower matter density observed by DESI are tightly constrained by Planck, which places stringent constraints on the allowed velocity kicks. The velocity kick scales as $v_{\rm kick} \propto (1-\varepsilon^2)^{1/2}$, so even small mass splittings produce significant velocities. We find the $1\sigma$ upper bound on the velocity kick ranging from  $\sim 10^{-2}c$ to $\sim 10^{-3}c$ when half of dark matter is unstable, with $v_{\rm kick} \lesssim 0.005c$ reached at the lifetime $\Gamma^{-1} \approx 1\,\mathrm{Gyr}$.

These kicks have several consequences. First, decay products suppress power on scales smaller than the free-streaming length. This in turn reduces the growth rate of the remaining stable CDM component making the clustering of matter less efficient. Finally, it can also alter the expansion history affecting the late-time evolution of the Universe. CMB lensing, which probes the integrated matter distribution, is particularly sensitive to these effects. 

Nonetheless, the allowed DDM parameter space remains of interest for small-scale structure anomalies. Our results also imply that solutions to the DESI tensions that involve decays to one massive and one massless particle \cite{Acharya:2026bmo} will be disfavored by CMB lensing.

The rest of this paper is organized as follows. In Sec.~\ref{sec:ddm-model}, we describe the DDM model. In Sec.~\ref{sec:fisher}, we describe the generalized Fisher analysis, and present our constraints on the velocity kick from Planck CMB, together with their implications in Sec.~\ref{sec:constraints}. We conclude in Sec.~\ref{sec:conclusion}.

\section{DDM Model}
\label{sec:ddm-model}
We consider a two-body decay in which a cold dark matter particle $\chi$ with mass $m_\chi$ decays into two identical decay products, each with mass $m_{\chi_{1,2}}=\varepsilon m_\chi / 2$.\footnote{The analysis can easily be generalized to unequal particle masses, and our code allows for this possibility.} The model is fully characterized by three parameters: the decay rate $\Gamma$, the fraction $f$ of the cold dark matter that is unstable, and the total mass retention $\varepsilon$. Our results do not depend on the mass of the decaying particle, only the ratio between its mass and the decay products' mass. Each decay product receives a velocity kick $v_\text{kick} = c(1-\varepsilon^2)^{1/2}$, and equivalently, momentum $p = (m_\chi c/2) (1 - \varepsilon^2)^{1/2}$.

We define a dimensionless comoving momentum  $q=a(\tau_q) p/T_0$, where $T_0$ is the current CMB temperature, $\tau_q$ is the conformal time at which the decay products are produced, and $a(\tau)$ is the scale factor, normalized to $a=1$ today. Each decay product thus has energy $\epsilon(q,\tau) = \left[ q^2 +a^2(\tau) m_{\chi_{1,2}}^2/T_0^2 \right]^{1/2}$ (in units of the CMB temperature).

The comoving freestreaming length of each decay product is 
\begin{equation}
\label{eq:lambdafs}
    \lambda_\text{fs} = \int_0^{z_d} \frac{v(z)}{H(z)}dz =  \frac{v_{\rm kick}}{1+z_d} \int_{0}^{z_d} \frac{1+z}{H(z)}dz,
\end{equation}
where $H(z)$ is the Hubble rate and $z_d$ is the characteristic decay redshift such that $\Gamma\approx H(z_d)$. In the second equality we approximate that the particle is nonrelativistic at decay, in which case $v(z)\propto 1+z$. If the particle decays during matter domination, $z_d \gg 1$, this becomes
\[
\lambda_{\rm fs} \approx 11{,}000\frac{v_{\rm kick}/c}{\sqrt{1+z_d}}\, h^{-1}\,\mathrm{Mpc},
\]
where $h = H_0 /(100\,$km\,s$^{-1}$\,Mpc$^{-1}$).
The relevant wavenumber and (primary CMB) multipole moment are $k_{\rm fs} = 2\pi/\lambda_{\rm fs}$ and $\ell_{\rm fs} \sim \pi \chi^*/\lambda_{\rm fs}$, respectively, where $\chi^* \approx 14{,}000\,$Mpc is the comoving distance to last scattering. 

\subsection{Background Evolution}
The energy density $\rho_\chi$ of the decaying particle evolves as
\begin{equation}
    \pdv{\rho_\chi}{\ln a} = -3\rho_\chi - \frac{\Gamma}{H}\rho_\chi.
\end{equation}
The phase-space density $f_0$ for the decay products satisfies \cite{Aoyama:2011, Aoyama:2014tga}
\begin{eqnarray}
     \frac{\partial f_0(q,\tau)}{\partial\tau} &=& \frac{a(\tau) \Gamma N_\chi(\tau)}{4 \pi q^2} \delta_D\left(q- \frac{a(\tau) p}{T_0} \right) \nonumber \\ 
     &= & \frac{a(\tau) \Gamma N_\chi(\tau)}{4 \pi q^3 {\cal H}(\tau)} \delta_D(\tau- \tau_q),
\label{eqn:uniformevolution}     
\end{eqnarray}
where $N_\chi$ is the comoving number density of the decaying particle, $\delta_D(x)$ is the Dirac delta function, and ${\cal H}\equiv(da/d\tau)/a$. The solution is then
\begin{eqnarray}
     f_0(q,\tau) &=&\widetilde{f}_0(q) \Theta(\tau-\tau_q),
\label{eq:f0}
\end{eqnarray}
where $\widetilde{f}_0(q)$ is the amplitude of the distribution determined at the time of decay $\tau_q$,
\begin{eqnarray}
    \widetilde{f}_0(q) &\equiv&  \frac{\Gamma a(\tau_q) N_\chi(\tau_q)}{4\pi q^3 {\cal H}(\tau_q)},
\label{eq:f0-tilde}
\end{eqnarray}
and $\Theta(x)$ is the Heaviside step function. The comoving average number density of these decay products is then $N(\tau) = \int\, d^3 q \;f_0(q,\tau)$.

\subsection{Perturbations}
With a perturbation of a Fourier mode with wavenumber $\vec{k}$, the phase-space distribution is written as $f(\vec{q},k,\tau) = f_0(q,\tau) + \Delta f(q,k,\mu,\tau)$. Here $\mu$ is the cosine of the angle between ${\vec k}$ and the direction of the particle momentum. The perturbation $\Delta f(q,k,\mu,\tau)$ satisfies \cite{Aoyama:2014tga}
\begin{eqnarray}
     \frac{\partial \Delta f}{\partial \tau} + ik\mu\frac{q}{\epsilon}\Delta f + \frac{\partial f_0}{\partial\ln q}\left[ \dot{\eta} - \frac{\dot{h}+6\dot{\eta}}{2}\mu^2\right] =  \frac{\partial f_0}{\partial \tau} \delta_\chi(k,\tau).\nonumber \\
\label{eq:Boltzmann}
\end{eqnarray}
where $\delta_\chi$ is the decaying particle density perturbation. The functions $h(k,\tau)$ and  $\eta(k,\tau)$ are the synchronous-gauge metric perturbations, and the dots denote derivative with respect to conformal time $\tau$. The solution is
\begin{eqnarray}
     \Delta f(\tau) &=& \int_0^\tau\, d\tau'\, e^{-i \mu k \chi_q(\tau',\tau)} \left\{   \frac{\partial f_0(q,\tau')}{\partial \tau} \delta_\chi(k,\tau) \right. \nonumber\\
     &&\left. +  \frac{ \partial f_0(q,\tau')}{\partial \ln q} \left[- \dot \eta (\tau') +\frac{\dot h(\tau') +6 \dot \eta(\tau')}{2}\mu^2  \right] \right\},~~~~~
\end{eqnarray}
where $\chi_q(\tau_1,\tau_2)\equiv \int_{\tau_1}^{\tau_2} d\tau' q/\epsilon(\tau')$ is the comoving distance traveled by the particle from $\tau_1$ to $\tau_2$.
The multipole moments $(\Delta f)_\ell \equiv (i^\ell/2)\int_{-1}^1 \Delta f(\mu) P_\ell(\mu)$, then become
\begin{eqnarray}
     (\Delta f)_l(\tau) &=&   \int_0^{\tau} d\tau'\; \Bigg\{  j_\ell[k \chi_q(\tau',\tau)] \frac{\partial f_0(q, \tau')}{\partial \tau} \delta_\chi(\tau') \nonumber \\
      &+& \frac{\partial f_0(q, \tau')}{\partial \ln q} \Bigg[   -j_\ell[k\chi_q(\tau',\tau)] \dot{\eta}  \nonumber \\
      & & - j''_\ell[k\chi_q(\tau',\tau)] \frac{\dot{h} + 6\dot{\eta}}{2} \Bigg]
   \Bigg\},
    \label{eq:Psi_ell}
\end{eqnarray}
where $j''_\ell(x) \equiv d^2 j_\ell(x) / dx^2$. 
The contributions to the stress-energy tensor from the perturbations of the decay products are
\begin{eqnarray}
\delta \rho &=& 4\pi a^{-4} T_0 \int q^2 dq\, \epsilon\, (\Delta f)_0(q), \nonumber \\
\delta P &=& \frac{4\pi}{3} a^{-4} T_0 \int q^2 dq\, \frac{q^2}{\epsilon}\, (\Delta f)_0(q), \nonumber \\
(\bar{\rho} + \bar{P})\theta
&=& 4\pi k a^{-4} T_0 \int q^2 dq\, q\, (\Delta f)_1(q), \nonumber \\
(\bar{\rho} + \bar{P})\sigma
&=& \frac{8\pi}{3} a^{-4} T_0 \int q^2 dq\, \frac{q^2}{\epsilon}\, (\Delta f)_2(q),
\label{eq:neutrino-perturb}
\end{eqnarray}
where the extra factor of $T_0$ accounts for the units of $f$.

We implement this DDM model in \texttt{CLASSIER} \cite{Ji:2022iji, Lee:2025zym, Lee:2025vgv}, employing the integral-equation approach for the decay product perturbations, resulting in \texttt{CLASSIER-DDM} \cite{Lee:2026zrl}. In this code, the Boltzmann hierarchies for the decay products are replaced by integral equations that are more efficiently computed. The details \nl{are} provided in Ref. \cite{Lee:2026zrl}. 

With the DDM model implemented in \texttt{CLASSIER-DDM}, we conducted best-fit analyses with DESI DR2 BAO measurements \cite{DESI:2025zgx} and the Planck 2018 CMB dataset \cite{Planck:2018vyg}. We found that the DDM model did not improve the fit compared to $\Lambda$CDM. We first provide physical intuition on the effects of DDM on CMB data before deriving constraints on the DDM model.

\subsection{Physical effects of DDM on CMB observables}
\label{sec:physical-effects}
In this Section, we describe the mechanisms through which DDM modifies the CMB observables. Each operates on different angular scales and depends differently on the DDM parameters.

\begin{figure}[h!]
    \centering
    \includegraphics[width=\linewidth]{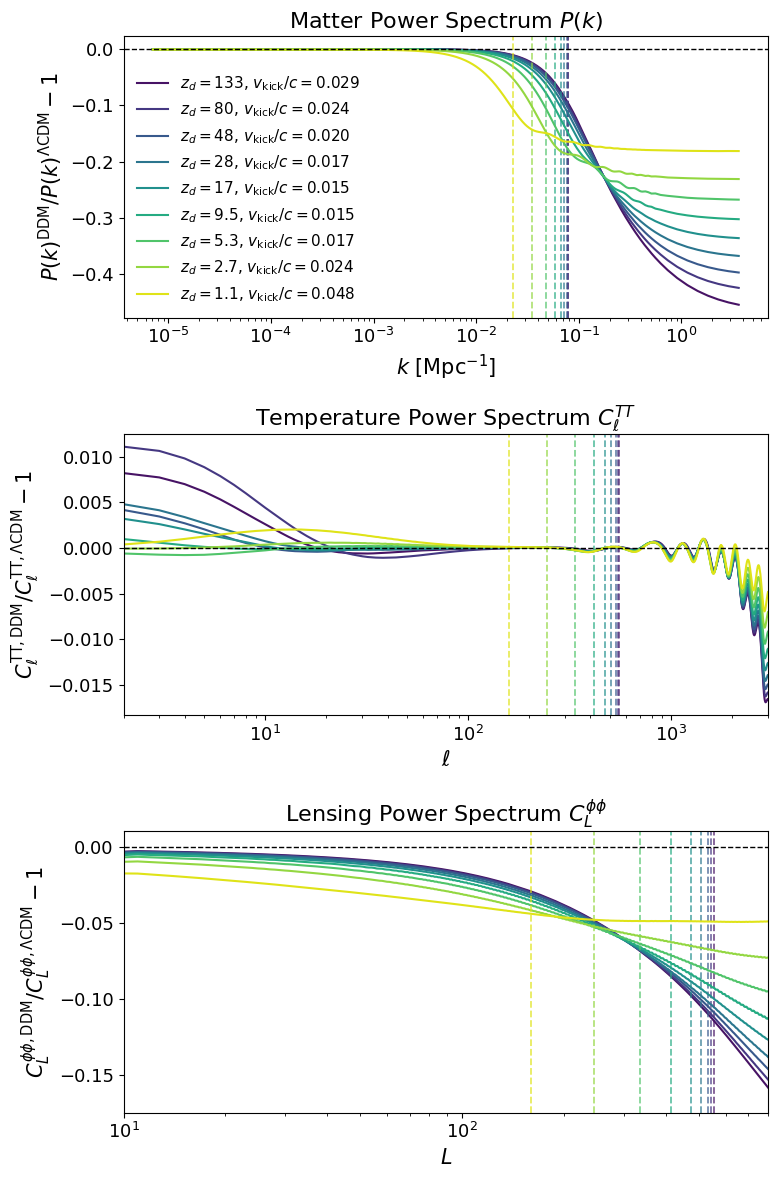} 
    \caption{Fractional deviations from $\Lambda$CDM in the matter power spectrum (top), temperature power spectrum $C^{TT}_\ell$ (middle), and lensing power spectrum $C^{\phi\phi}_L$ (bottom). Each curve corresponds to a DDM model with a different decay lifetime $\Gamma^{-1}$ at fixed $f = 0.1$, with the mass retention parameter $\varepsilon$ set such that it yields the 1$\sigma$ upper limit on $v_{\rm kick}$. The vertical dashed lines in the $P(k)$ and $C_\ell$ panels indicate the free-streaming wavenumber $k_{\rm fs} = 2\pi/\lambda_{\rm fs}$ and free-streaming multipole moment $\ell_{\rm fs}$, respectively, for selected models.}
    \label{fig:threshold-spectra}
\end{figure}

In Fig.~\ref{fig:threshold-spectra}, we show fractional deviations from $\Lambda$CDM in the matter power spectrum (top), temperature power spectrum (middle), and lensing power spectrum (bottom). Each curve corresponds to a different decay lifetime $\Gamma^{-1}$, with $f=0.1$ fixed. For each $\Gamma^{-1}$, the mass parameter $\varepsilon$ is chosen such that $v_{\rm kick}$ sits at the 1$\sigma$ upper limit derived in Sec~\ref{sec:fisher}, so that each curve illustrates a deviation from $\Lambda$CDM at the $1\sigma$ level. 

After decay, the decay products receive a velocity kick $v_{\rm kick} = c(1-\varepsilon^2)^{1/2}$ in the parent rest frame. These particles free-stream across gravitational potential wells, smoothing out perturbations on scales smaller than the comoving free-streaming length $\lambda_{\rm fs}$. The resulting suppression in the matter power spectrum at $k \gtrsim \lambda_{\rm fs}^{-1}$ is visible in the top panel of Fig.~\ref{fig:threshold-spectra}, where the dashed lines mark $k_{\rm fs} = 2\pi/\lambda_{\rm fs}$.

The net suppression exceeds the naive expectation of a suppression proportional to $f$. Once the decay products free-stream out of small-scale potential wells, the growth rate of the remaining stable CDM component is reduced. This amplifies the overall power deficit, and explains why only a 10\% unstable fraction produces the large suppression visible in the top panel of Fig.~\ref{fig:threshold-spectra}. It also explains why earlier decays lead to stronger suppression, since this effect accumulates throughout time.\footnote{The small-scale suppression from massive neutrinos is significantly weaker than that from the decay products. For the parameters considered here (one massive neutrino species with $mn_\nu = 0.06$ eV), the difference between a zero and non-zero neutrino mass amounts to $\sim2\%$ suppression in $P(k)$ at $k\gtrsim0.1\,\rm{Mpc}^{-1}$, compared with the $\sim40\%$ suppression from the decay. The lensing effect is purely due to decays.}

This small-scale power suppression propagates directly into the CMB lensing potential as the reduced matter clustering along the line of sight produce a weaker lensing signal. The range of multipoles at which the CMB lensing power spectrum $C_L^{\phi\phi}$ is suppressed directly reflects the range of $k$ at which $P(k)$ is suppressed. The bottom panel of Fig.~\ref{fig:threshold-spectra} shows that the fractional deviation in $C_L^{\phi\phi}$ reaches $\sim$ 5--10\% within the multipole range $L\leq400$ covered by Planck's conservative lensing reconstruction \cite{Planck:2018lbu}, which is comparable to the measurement uncertainties. 

Finally, in $\Lambda$CDM, gravitational potentials remain nearly constant during matter domination, producing a minimal integrated Sachs-Wolfe (ISW) contribution to the CMB temperature anisotropies. There is a departure from matter domination in DDM shortly after decay when the decay products are warm, but this effect is very small for the parameter values we are dealing with. The changes seen in the middle panel of Fig.~\ref{fig:threshold-spectra} at low $\ell$ arise from the small changes to the matter density at low redshift induced by DDM, consistent with similar enhancements seen in decays to one massive and one massless particle \cite{Aoyama:2011,Aoyama:2014tga,Abell_n_2021}.  

\section{Generalized Fisher analysis}
\label{sec:fisher}

We may now derive constraints on the velocity kick $v_{\rm kick} = c(1-\varepsilon^2)^{1/2}$ from Planck 2018 CMB data. To this end, we apply the generalized Fisher analysis of Ref.~\cite{Lee:2021bmn}, which accommodates the non-Gaussian dependence of the likelihood on $v_{\rm kick}$. This generalization is necessary because the changes in the CMB spectra induced by DDM depend on $v_{\rm kick}^2$ to leading order,\footnote{The quantity $v_{\rm kick}$ enters into the arguments of the spherical Bessel functions in Eq.~(\ref{eq:Psi_ell}).  The derivatives of those spherical Bessel functions with respect to $v_{\rm kick}$ are quadratic in $v_{\rm kick}$ for $\ell=0$ and 2, but linear for $\ell=1$.  However, $(\Delta f)_{\ell=1}(\tau)$ then enters in the expression for $\theta$ in Eq.~\eqref{eq:neutrino-perturb} which has an additional factor of $q$ (also proportional to $v_{\rm kick}$) relative to the expression for $\delta\rho$.} so the likelihood is not Gaussian in $v_{\rm kick}$, and the fiducial value $v_{\rm kick} = 0$ lies at the boundary of the physical parameter space, making a standard Fisher analysis inapplicable. For the analysis, we consider a grid over $(f,\, \Gamma^{-1})$ with $0.1 \leq f  \leq 0.5$ and $0.01 \lesssim \Gamma^{-1} \lesssim 10$ Gyr. At each grid point, we compute the CMB angular power spectra $\boldsymbol{C} \equiv \{C_\ell^{\rm TT}, C_\ell^{\rm TE}, C_\ell^{\rm EE}, C_L^{\phi\phi}\}$ using \texttt{CLASSIER-DDM}.\footnote{We restrict to $f\leq0.5$ as the integral-equation approach becomes computationally expensive for large $f$, where the decay products contribute significantly to the total energy. We defer a dedicated treatment of this regime to future work.}

The method proceeds as follows. We denote the standard $\Lambda$CDM parameters by $\boldsymbol{p} \equiv \{\omega_\Lambda, \omega_c, \omega_b, \tau_{\rm reio}, \ln(10^{10}A_s), n_s\}$, with fiducial values $\boldsymbol{p}^{\rm fid}$ taken to be the Planck 2018 best-fit. Let $\boldsymbol{M}$ denote the inverse of the data covariance matrix. The $\chi^2$ is
\begin{equation}
\chi^2 = \left(\boldsymbol{C}(v_{\rm kick}, \boldsymbol{p}) - \boldsymbol{C}^{\rm obs}\right) \cdot \boldsymbol{M} \cdot \left(\boldsymbol{C}(v_{\rm kick}, \boldsymbol{p}) - \boldsymbol{C}^{\rm obs}\right),
\end{equation}
where $\boldsymbol{C}^{\rm obs}$ is the observed data. We approximate $\boldsymbol{C}^{\rm obs} \approx \boldsymbol{C}(0, \boldsymbol{p}^{\rm fid})$, replacing the data with the theory prediction at the fiducial $\Lambda$CDM parameters. This drops the residual $\boldsymbol{C}(0, \boldsymbol{p}^{\rm fid}) - \boldsymbol{C}^{\rm obs}$ and the cross term between this residual and $\boldsymbol{C}(v_{\rm kick}, \boldsymbol{p}) - \boldsymbol{C}(0, \boldsymbol{p}^{\rm fid})$, both of which are small since $\boldsymbol{p}^{\rm fid}$ is a good fit to the data. The approximated $\chi^2$ then depends only on the change in spectra,
\begin{equation}
\Delta\boldsymbol{C}_{\rm DDM}(v_{\rm kick}) \equiv \boldsymbol{C}(v_{\rm kick},\, \boldsymbol{p}^{\rm fid}) - \boldsymbol{C}(0,\, \boldsymbol{p}^{\rm fid}),
\end{equation}
computed by direct evaluation at a grid of $v_{\rm kick}$ values. We decompose $\Delta\boldsymbol{C}_{\rm DDM}$ into a part degenerate with $\Lambda$CDM parameter variations and a perpendicular remainder:
\begin{equation}
\Delta\boldsymbol{C}_{\rm DDM} = \sum_{i=1}^{6} \alpha_i\, \frac{\partial \boldsymbol{C}}{\partial p_i} + \Delta\boldsymbol{C}^{\perp}_{\rm DDM},
\label{eq:decomp}
\end{equation}
where orthogonality is defined with respect to $\boldsymbol{M}$ as the inner product,
\begin{equation}
\frac{\partial \boldsymbol{C}}{\partial p_j} \cdot \boldsymbol{M} \cdot \Delta\boldsymbol{C}^{\perp}_{\rm DDM} = 0, \qquad \forall\; j = 1, \ldots, 6.
\end{equation}
The coefficients $\alpha_i$ are obtained by solving
\begin{equation}
\alpha_i = \sum_{j=1}^{6} \left(F^{-1}\right)_{ij} \frac{\partial \boldsymbol{C}}{\partial p_j} \cdot \boldsymbol{M} \cdot \Delta\boldsymbol{C}_{\rm DDM},
\end{equation}
where
\begin{equation}
F_{ij} \equiv \frac{\partial \boldsymbol{C}}{\partial p_i} \cdot \boldsymbol{M} \cdot \frac{\partial \boldsymbol{C}}{\partial p_j}, \qquad 1 \leq i,j \leq 6,
\end{equation}
is the $6\times 6$ Fisher matrix of the standard $\Lambda$CDM parameters. Integrating the full likelihood over $\boldsymbol{p}$, the marginalized likelihood for $v_{\rm kick}$ is
\begin{equation}
\mathcal{L}(v_{\rm kick}) \propto \exp\!\left(-\frac{1}{2}\,\Delta\boldsymbol{C}^{\perp}_{\rm DDM}\cdot \boldsymbol{M} \cdot\Delta\boldsymbol{C}^{\perp}_{\rm DDM}\right).
\label{eq:marg_like}
\end{equation}
We estimate the 1$\sigma$ upper limit on $v_{\rm kick}$by finding the value at which
\begin{equation}
\Delta\boldsymbol{C}^{\perp}_{\rm DDM} \cdot \boldsymbol{M} \cdot \Delta\boldsymbol{C}^{\perp}_{\rm DDM} = 1.
\label{eq:1sigma}
\end{equation}
Since $\Delta\boldsymbol{C}^{\perp}_{\rm DDM}$ depends on $v_{\rm kick}$ in a way that is generally not linear --- and whose precise form depends on $\Gamma$ --- the marginalized posterior is not Gaussian in $v_{\rm kick}$, and Eq.~\eqref{eq:1sigma} is, strictly speaking, only exact for a Gaussian posterior. Nevertheless, this criterion provides a useful estimate across the parameter space. When the posterior is broad, its precise shape matters little, and when the allowed region is narrow, Eq.~\eqref{eq:1sigma} remains a reasonable approximation to the upper limit. The inverse covariance matrix $\boldsymbol{M}$ comprises the compressed low-$\ell$ TT and EE likelihood of Refs.~\cite{Lee:2026toh, Prince:2021fdv}\footnote{Available at \href{https://github.com/nanoomlee/planck-gaussian-lowl}{github.com/nanoomlee/planck-gaussian-lowl}.}, the high-$\ell$ (TT, TE, EE) Planck-lite likelihood~\cite{Planck:2018vyg}, and the CMB lensing measurements~\cite{Planck:2018lbu} accessed via \texttt{Cobaya}~\cite{Torrado:2020dgo}. 

\section{CMB Constraints on DDM}
\label{sec:constraints}
In this Section, we use the generalized Fisher analysis described above to derive CMB constraints on the velocity kick received by the decay products. 

\subsection{Quantitative constraints on the velocity kick}
\label{sec:constraints-vkick}

\begin{figure}[t]
    \centering
    \includegraphics[width=\linewidth]{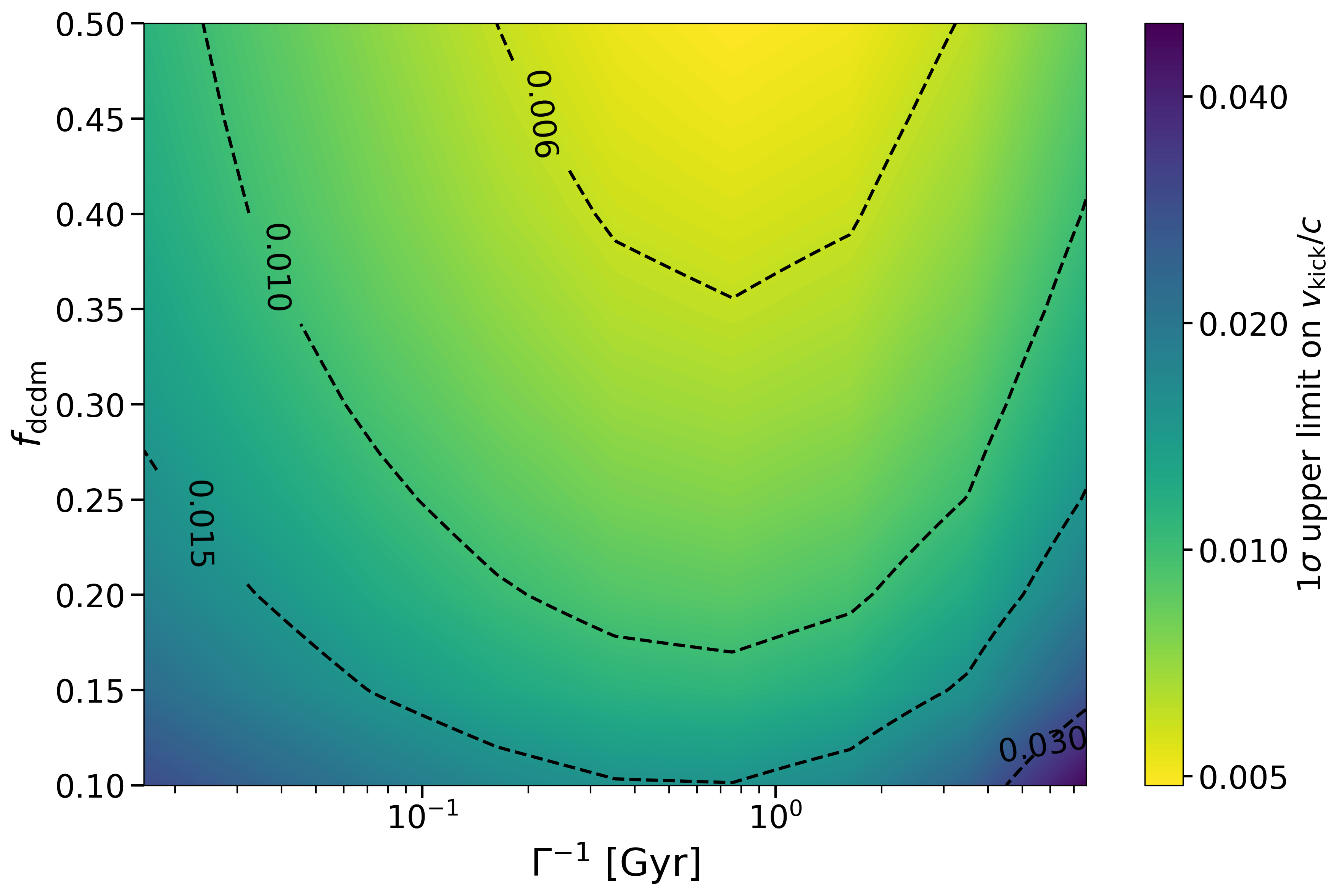}
    \caption{The $1\sigma$ upper limit on $(v_{\rm kick}/c)$ in the $(\Gamma^{-1}, f)$ plane. Yellow regions indicate tighter constraints (smaller allowed deviations from $v_{\rm kick} = 0$).}
    \label{fig:fisher-exclusion}
\end{figure}

\begin{figure*}[t]
    \centering
    \includegraphics[width=\linewidth]{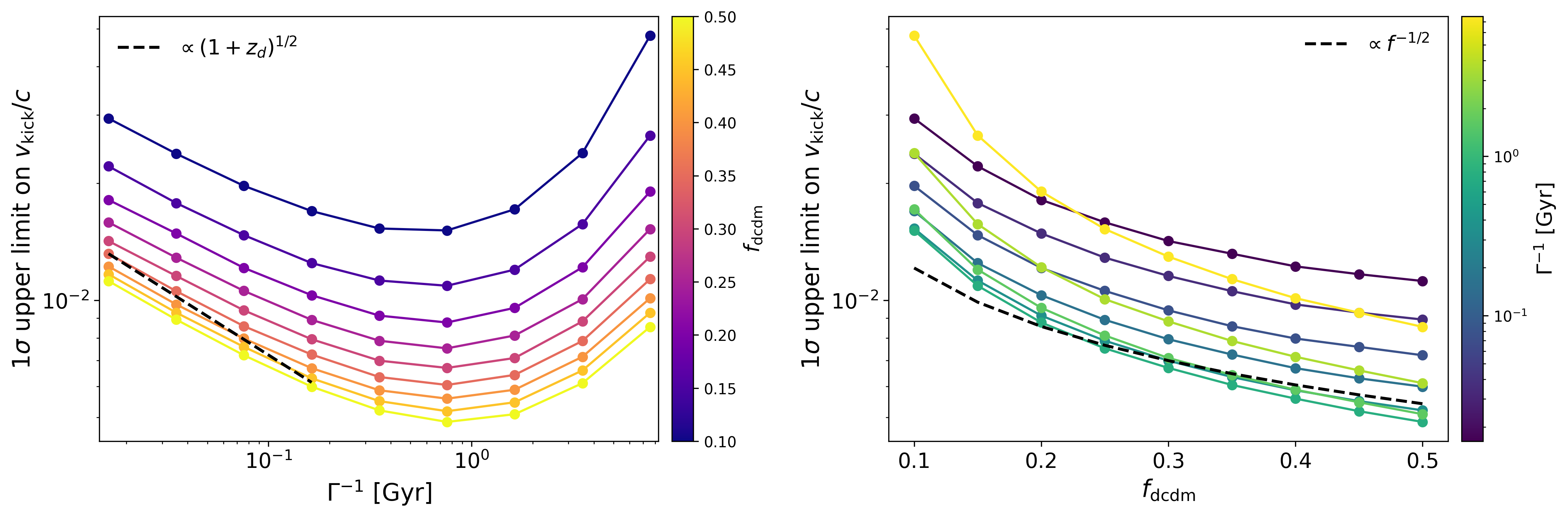}
    \caption{The $1\sigma$ upper limit to $v_{\rm kick}$ in units of the speed of light as a function of DDM parameters. \textit{Left:} 1$\sigma$ upper limit on $v_{\rm kick}/c$ versus $\Gamma^{-1}$, with curves colored by $f$. The black dashed line shows the expected scaling $\propto (1+z_d)^{1/2}$ for $\Gamma^{-1} < 1\,$Gyr. \textit{Right:} 1$\sigma$ upper limit on $v_{\rm kick}/c$ versus $f$, with curves colored by $\Gamma^{-1}$. The black dashed line shows the expected scaling $\propto f^{-1/2}$.}
    \label{fig:fisher-vs-f-gamma}
\end{figure*}

In Fig.~\ref{fig:fisher-exclusion}, we show the marginalized 1$\sigma$ upper limit on $v_{\rm kick}$ across the $(\Gamma^{-1},\,f)$ plane. The bounds are most restrictive at intermediate lifetimes $\Gamma^{-1} \sim 1~$Gyr, with the $1\sigma$ upper limit reaching $v_{\rm kick} \lesssim 0.005c$ for the unstable fraction $f=0.5$. We find that the $\chi^2$ contribution is dominated by the lensing data, indicating that CMB lensing provides the dominant constraining power in our analysis.

The dependence of the upper bound on $\Gamma^{-1}$ and $f$ can be understood from the physical effects described in Section~\ref{sec:physical-effects}. Figure~\ref{fig:fisher-vs-f-gamma} (left) shows the 1$\sigma$ bound on $v_{\rm kick}$ as a function of $\Gamma^{-1}$. At short lifetimes ($\Gamma^{-1} \lesssim 0.1\,$Gyr), decays occur during earlier periods of matter domination, and the decay products are decelerated more rapidly, reducing the comoving free-streaming length as $(1+z_d)^{-1/2}$, Eq.~\eqref{eq:lambdafs}. The free-streaming suppression is therefore pushed to scales smaller than those probed by CMB lensing, and the upper bounds on $v_{\rm kick}$ correspondingly increase. At long lifetimes, a large fraction of the unstable component has not yet decayed by today, reducing the observable signal and widening the allowed region. 

Figure~\ref{fig:fisher-vs-f-gamma} (right) shows the 1$\sigma$ bound as a function of $f$. The constraint tightens with increasing $f$, as a larger unstable fraction produces a stronger signal. At leading order, since the observable signal in the perturbations scales as $f\cdot v_{\rm kick}^2$, one would expect the $1\sigma$ upper limit to scale as $f^{-1/2}$, shown as the black dashed line. The derived bounds are broadly consistent with this scaling.

\subsection{Implications for the DDM parameter space}

These constraints have direct implications for two classes of DDM models: those proposed to resolve cosmological tensions, and those designed to address small-scale structure anomalies. 

First, for the DDM parameter spaces we considered, translating the 1$\sigma$ upper limit on $v_{\rm kick}$ into the corresponding fractional mass loss limits it to $\Delta \Omega_m / \Omega_m \lesssim 0.1\%$. For comparison, resolving the apparent discrepancy between the matter density inferred from Planck and that preferred by DESI BAO~\cite{Planck:2018lbu, DESI:2025zgx} would require $\Delta \Omega_m/\Omega_m \sim 5\%$, corresponding to $v_{\rm kick} \gtrsim 0.5c$, which is orders of magnitude outside of our constraints. 

These bounds translate directly into constraints on the fractional mass difference between the decaying particle and its decay products, $\Delta m / m$. For decays to two massive particles, the kick velocity scales as $v_{\rm kick} \simeq c(2\Delta m/m)^{1/2}$ for $\Delta m /m \ll 1$; our bounds then require $(\Delta m/m) \lesssim 10^{-3}$. For decays to one massive and one massless decay product, the kinematics instead give $v_{\rm kick} \simeq c(\Delta m/m)$, yielding $(\Delta m/m) \lesssim 10^{-2}$. Both cases are in broad agreement with earlier work~\cite{Aoyama:2011, Aoyama:2014tga, Abell_n_2021, Bucko:2022kss}. The fractional mass splitting $(\Delta m/m) \sim 0.05$--$0.1$ required to account for the DESI discrepancy exceeds our bounds in either scenario.

Second, DDM models addressing small-scale anomalies require free-streaming on sub-galactic scales, $\lambda_{\rm fs} \sim 1\,h^{-1}\,\mathrm{Mpc}$. Our CMB constraints are consistent with this regime; models with $\lambda_{\rm fs} \sim 1$--$10\, h^{-1}\,\mathrm{Mpc}$ and $v_{\rm kick} \lesssim 500\,\mathrm{km\,s^{-1}}$ are consistent with our bounds while producing suppression on the sub-galactic scales relevant to dwarf-galaxy anomalies~\cite{Bullock:2017xww, Abdelqader:2008wa, Ando:2021fhj}.  It will also be interesting to consider the impacts of these models on the Lyman-alpha-forest power spectrum \cite{Fuss:2022zyt,Zhao:2026wxi,Wang:2013rha}.

\section{Conclusion}
\label{sec:conclusion}

We have derived constraints from Planck 2018 CMB data on dark matter that decays to two massive particles using \texttt{CLASSIER-DDM} \cite{Lee:2026zrl,Ji:2022iji, Lee:2025zym, Lee:2025vgv}, a modified version of the public Boltzmann code \texttt{CLASS} that efficiently handles warm decay products via an integral-equation method.

DDM does not improve the fit to either DESI BAO or Planck CMB data relative to $\Lambda$CDM. Our generalized Fisher analysis yields a marginalized $1\sigma$ upper bound ranging from  $\sim 10^{-2}c$ to $\sim 10^{-3}c$ when half of dark matter is unstable (Fig.~\ref{fig:fisher-exclusion}), with $v_{\rm kick} \lesssim 0.005c$ reached at the lifetime $\Gamma^{-1} = 1~\mathrm{Gyr}$. The 1$\sigma$ upper limits we derive roughly scale as $\propto f^{-1/2}$ and $\propto (1+z_d)^{1/2}$ for large $z_d$ (Fig.~\ref{fig:fisher-vs-f-gamma}). Extrapolating to fully unstable dark matter ($f=1$), our constraint projects to $v_{\rm kick} \lesssim 0.003c$.  

Throughout our analysis, CMB lensing provides the dominant constraint: models at the $1\sigma$ boundary produce 5--10\% deviations in the lensing power spectrum over the multipole range probed by Planck. The free-streaming lengths at these boundaries range from $\lambda_{\rm fs} \sim 30 -200\,h^{-1}\,\mathrm{Mpc}$, depending on $f$ and $\Gamma^{-1}$. DDM models addressing dwarf-galaxy anomalies require free-streaming on sub-galactic scales, $\lambda_{\rm fs} \lesssim \mathrm{few}\,h^{-1}\,\mathrm{Mpc}$, which fall well below the reach of CMB lensing. These models are therefore unconstrained by the present analysis. Complementary probes such as strong lensing are required to test them.

A full MCMC exploration would provide proper posterior distributions and enable marginalization over the complete parameter space. Combining CMB data with weak lensing surveys (e.g., DES, HSC, future Rubin Observatory data) would narrow the allowed region of the matter power spectrum at $k \sim 0.1$--$1\,h\,\mathrm{Mpc}^{-1}$ \cite{DES:2017myr, Sugiyama:2023fzm, LSSTDarkMatterGroup:2019mwo}. As was shown in \cite{Gilman:2026uvq}, strong lensing also provides significant constraints on decaying dark matter, but may benefit from a more careful treatment of the transfer function. Including ACT data would extend the analysis to smaller angular scales with greater sensitivity to lensing, further narrowing the allowed parameter space. Finally, N-body simulations of DDM cosmologies could predict observable consequences for halo mass functions and satellite abundances, connecting our constraints to small-scale structure observations.

\begin{acknowledgments}
We thank Adrienne Erickcek for useful discussions. This work was supported at JHU by NSF Grant No.\ 2412361, NASA ATP Grant No.\ 80NSSC24K1226, and the Templeton Foundation. NL was supported by the Horizon Fellowship from Johns Hopkins University.
\end{acknowledgments}

\bibliography{bib}
\end{document}